\documentclass[preprints,accept,pdftex,moreauthors]{mdpi} 
\usepackage{amsmath}
\usepackage{amsfonts}
\usepackage{amssymb}
\usepackage{euscript}
\firstpage{1} 
\pubvolume{1}
\issuenum{1}
\articlenumber{0}
\pubyear{2026}
\copyrightyear{2026}
\datereceived{ } 
\daterevised{ } 
\dateaccepted{ } 
\datepublished{ } 

\def\prn#1{{\left(#1\right)}}

\def\abs#1{{\left|#1\right|}}

\Title{Nuclear Spin Oscillator Based on $^3$He to Search for Exotic Spin Coupling }

\Author{Heather R. Pearson$^\dagger$, Anna Molodtsova$^\ddagger$, Sage C.\  Weisrock$^{\S}$, Sherlock Tingrui Zhao$^{\|}$, and Jason E.\ Stalnaker *}

\AuthorNames{Firstname Lastname, Firstname Lastname and Firstname Lastname}

\address{\quad Department of Physics and Astronomy, Oberlin College; Oberlin, OH 44074, USA}

\corres{Correspondence: jason.stalnaker@oberlin.edu}

\firstnote{Current address: Department of Physics and Astronomy, University of California, Davis; Davis, CA 95616, USA.}  
\secondnote{Current address: Department of Mechanical Engineering, Columbia University; New York, NY 10027, USA}
\thirdnote{Current address: Department of Physics and Astronomy, University of New Mexico; Albuquerque, NM 87106, USA}
\fourthnote{Current address: Department of Physics, Yale University; New Haven, CT 06520, USA}

\abstract{We describe an experimental investigation of a nuclear spin oscillator based on $^3$He nuclei as a possible detector to search for exotic spin couplings. A magnetically shielded vapor cell comprised of an alkali atom mixture ($95\%$ potassium and $5\%$ rubidium) and $^3$He gas is polarized via laser light resonant with the $D_1$ transition in rubidium in the presence of a dc magnetic field. The potassium atoms and $^3$He nuclei are polarized via spin-exchange collisions with the polarized rubidium atoms. The nuclear spins are tipped with a magnetic field applied perpendicular to the dc magnetic field. The resulting Larmor precession of the $^3$He nuclear spins is monitored via Faraday rotation of laser light near resonant with the $D_1$ transition in potassium. The Faraday rotation signal is filtered, amplified, and used to apply a magnetic field in a direction perpendicular to the dc magnetic field, resulting in a self-sustained oscillation of the nuclear spins at a frequency that is directly proportional to the dc magnetic field. We demonstrate a sensitivity to exotic spin couplings that is $\approx 5$ times higher than the alkali atom magnetometers that have been used in the Global Network of Optical Magnetometers to Search for Exotic Physics collaboration.
}

\keyword{nuclear spin oscillator; exotic spin coupling; magnetometry} 

\begin{document}


\section{Introduction}

The direct detection of dark matter remains elusive. Multiple hypothesized dark matter candidates have led to many direct detection experiments utilizing different experimental techniques (see, e.g.\ Reference \cite{budker23}, for a review of dark matter). The possibility that dark matter might be comprised of ultralight, bosonic, pseudoscalar particles has driven searches for dark matter that use optically pumped atomic magnetometers. Dark matter of this type may form structures larger than the Earth and behave as coherent fields rather than individual particles \cite{jacksonKimball23}. This has led to the development of \textbf{G}lobal \textbf{N}etwork of \textbf{O}ptical \textbf{M}agnetometers to Search for \textbf{E}xotic Physics (GNOME) collaboration. The collaboration consists of approximately fifteen optically pumped atomic magnetometers spread throughout the globe that look for transient signals due to interactions between the dark matter field and Standard Model particle spins. For an overview of the GNOME experiment and discussion of the dark matter candidates the network is targeting, see Reference \cite{afach23}. 

The GNOME collaboration has conducted multiple science runs using alkali-atom based magnetometers and placed limits on dark matter in the form of axion domain walls \cite{afach21} and solar relaxion halos \cite{wilson26}. In addition, the GNOME collaboration has placed constraints on exotic ultralight fields that may be produced during black hole mergers \cite{khamis25}.  Further analysis efforts looking for different dark matter candidates are ongoing.

Noble gas based magnetometers offer distinct advantages over alkali atom based magnetometers, primarily due to the higher number density achievable and finer energy resolution. As a result, many of the GNOME stations have upgraded or are in the process of upgrading the detectors to noble gas based magnetometers \cite{padniuk22,klinger23}. Here we present the results of an investigation into a noble gas magnetometer based on a nuclear spin oscillator with artificial feedback, sometimes called a nuclear spin maser, as a detector for dark matter. Nuclear spin oscillators of this type that are based on xenon nuclei have been suggested as detectors for searches for the permanent electric dipole moment \cite{yoshimi02,inoue13,inoue16,sato15,sato18} and oscillating dark matter fields \cite{jiang21}. The nuclear spin oscillator considered here differs from previous investigations in that it uses $^3$He spins. This system exhibits significantly longer transverse relaxation times than the nuclear spin oscillators based on xenon, making it more sensitive to magnetic fields or exotic spin couplings.

\section{Theory}

Nuclear spin oscillators rely on the Larmor precession of the nuclear spins of a noble gas in a dc magnetic field. The nuclear spins are polarized via spin-exchange collisions with optically pumped alkali atoms that are co-located with the noble gas. The precession of the nuclear spins is measured via Faraday rotation induced by the nuclear polarization on the alkali vapor. The Faraday rotation signal is amplified, phase shifted and fed back to magnetic field coils that apply a field in a perpendicular direction to the dc magnetic field. This feedback sustains the oscillation, creating a stable sinusoidal signal the frequency, $f$, of which is proportional to the magnetic field, $B$,
\begin{align}
  f = \gamma\, B\, , \label{eq:FGammaB} 
\end{align}
where $\gamma = 32.434 \:\textrm{Hz}/\mu\textrm{T}$ is the gyromagnetic ratio of the $^3$He atom \cite{mohr25}. One advantage of the nuclear spin oscillator as a magnetometer is this direct relationship between the frequency and the magnetic field, meaning no calibration of the magnetometer's response is required. \footnote{We note that alkali-noble gas magnetometers can be operated in a regime where the nuclear spins and alkali spins dynamically couple and the time evolution of the nuclear spins is not described by the simple Larmor precession frequency relation \cite{kornack02}.  This system operates far from this regime and the applied magnetic field is much larger than the field from the nuclear spins.} 

The behavior of a nuclear spin oscillator is well modeled by the Bloch equations. We consider a system of spins that are being polarized along the $\hat{z}$ axis in the presence of a leading field $\vec{B}_0 = B_0\, \hat{z}$. The net polarization vector of the spins, $\vec{P}$, in a magnetic field $\vec{B}$ is described by
\begin{align}
  \frac{d\vec{P}}{dt} = \gamma \vec{P}\times \vec{B} - \sum_{i=x,y,z}\Gamma_i \, P_i \, \hat{x}_i+ (P_\textrm{A} - P_z)\, G \, \hat{z}  ,
\end{align}
where $\gamma$ is the gyromagnetic ratio of the nuclear spins, $\Gamma_i$ describes the relaxation of the Cartesian components of the polarization, $\hat{x}_i$ are the unit Cartesian coordinate vectors, $P_\textrm{A}$ is the maximum polarization of the alkali-atom spins along the $\hat{z}$ axis, and $G$ is the pumping factor. In terms of the Cartesian vector components we have
\begin{align}
    \frac{dP_x}{dt} & = \gamma(P_y\, B_0 - P_z\, B_y) - \frac{P_x}{T_2} \\
    \frac{dP_y}{dt} & = \gamma(P_z\, B_x - P_x\, B_0) - \frac{P_y}{T_2} \\
    \frac{dP_z}{dt} &= \gamma(P_x\, B_y - P_y\, B_x) + \frac{P_0 - P_z}{T_1^*}, 
\end{align}
where $T_2$ is the transverse relaxation time and we have combined the longitudinal relaxation time, $T_1$ wuth the pumping rate of the nuclear spins, $G$, to get an effective longitudinal relaxation rate $T_1^* = T_1/\prn{1+T_1\, G}$ \cite{inoue16}. The equilibrium nuclear-spin polarization is then given by 
\begin{align}
  P_0 = P_\textrm{A} \, G\, T_1^* .
\end{align}
For the system considered here, $P_x$ is measured via Faraday rotation and that signal is fed back to the magnetic field coils that produce a magnetic field in $B_y$,
\begin{align}
  B_y\prn{t} = \alpha\, P_x\prn{t} ,
\end{align}
where $\alpha$ is the external feedback parameter. This provides a set of coupled differential equations 
\begin{align}
    \frac{dP_x}{dt} & = \gamma(P_y\, B_0 - \alpha\,P_z\, P_x ) - \frac{P_x}{T_2} \label{eq:NSODiffEqA} \\
    \frac{dP_y}{dt} & = - \gamma \, P_x\, B_0 - \frac{P_y}{T_2}  \label{eq:NSODiffEqB} \\ 
    \frac{dP_z}{dt} &= \alpha\, \gamma\, P_x^2  + \frac{P_0 - P_z}{T_1^*}, \label{eq:NSODiffEqC} 
\end{align}
where we have set $B_x = 0$. If the feedback signal and pumping rate are large enough to compensate the relaxation, the system will precess in a steady state operation. 

\section{Materials and Methods}

\begin{figure}[H]
\includegraphics[width=13.0 cm]{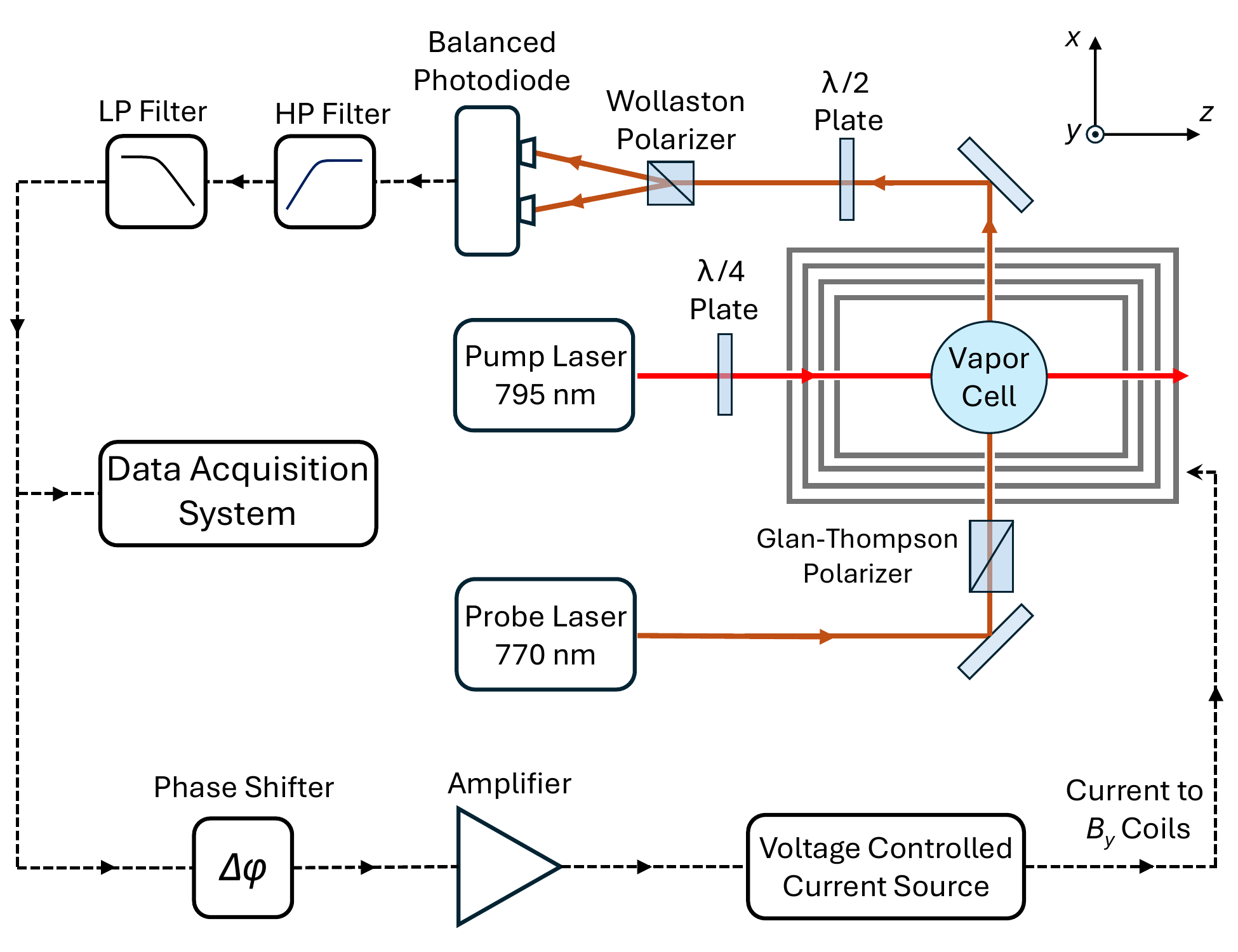}
\caption{Block diagram of the apparatus. Laser light resonant with the $D_1$ transition in rubidium (795~nm) polarized rubidium atoms contained in a vapor cell housed in a nested set of magnetic shields. The potassium atoms and $^3$He nuclei were polarized via spin-exchange collisions with the rubidium atoms. The component of the polarization of the $^3$He atoms along the $x$ direction was monitored via Faraday rotation of the probe laser light near the $D_1$ transition in potassium (770~nm) using a Wollaston polarizer and balanced photodiode. The Faraday rotation signal was filtered, phase shifted, amplified and fed back into a voltage controlled current source that produced a magnetic field in the $y$ direction. The filtered signal from the balanced photodiode was recorded by the data acquisition system. Not shown are the optics for the stabilization of both the pump and probe laser powers.  \label{fig:BlockDiagram}}
\end{figure}

The experimental setup is shown in Figure \ref{fig:BlockDiagram}. 

The vapor cell used in this experiment was a spherical cell with a diameter of 2.0~cm containing 3~amagat of $^3$He, 200~Torr of N$_2$, and a mixture of  $95\%$ K and 5$\%$ Rb. The cell was heated to 160 $^\circ$C and the densities of the potassium and rubidium atoms were measured via resonant absorption on the $D_1$ transitions to be $n_\textrm{K} = 2.0 \times 10^{13}\: \textrm{cm}^{-3}$ and $n_\textrm{Rb} = 1.0 \times 10^{12}\: \textrm{cm}^{-3}$, respectively. The composition of the rubidium and potassium was chosen so that the system could be optically pumped on rubidium and probed on potassium. This hybrid pumping scheme provides more efficient polarization of the $^3$He nuclei \cite{babcock03}. Potassium was chosen as the majority species over rubidium due to its lower spin-destruction cross section with $^3$He \cite{allred02}. The nitrogen gas was included to collisionally quench the alkali atoms and reduce radiation trapping.

The vapor cell was housed within four nested magnetic shields. The shielding reduces the effects of ambient magnetic fields by a factor of 10$^6$.  Inside the shields, magnetic field coils applied magnetic fields along $B_x$, $B_y$, and $B_z$, and magnetic field gradients $\frac{dB_y}{dx}$, $\frac{dB_y}{dy}$,  $\frac{dB_z}{dx}$, $\frac{dB_z}{dy}$, $\frac{dB_z}{dz}$, and $\frac{d^2B_z}{dz^2}$. These coils were used to zero out the magnetic fields and first-order gradients due to residual magnetization and apply the desired fields for the operation of the oscillator. The fields were zeroed with an applied leading field $B_z$ that was at or near the field used for the operation of the nuclear spin oscillator in order to account for possible gradients arising from imperfect polarization of the nuclear spins in the spherical cell. There was an additional smaller set of magnetic field coils inside the shields that were used to apply magnetic fields to characterize the response of the oscillator. 

A tapered amplifier, seeded with a distributed feedback laser, generated light resonant with the $D_1$ transition at 795~nm.  A piece of quartz picked off a portion of the light, which was monitored with a photodiode. This signal was used to stabilize the power by feeding the photodiode signal back to the current of the tapered amplifier. The light was circularly polarized and $\approx 550$~mW of power was used to optically pump the rubidium spins along the $z$ axis in the presence of a magnetic field of $\approx 300$ nT applied in the $+\hat{z}$ direction (see Figure \ref{fig:BlockDiagram}). The potassium atoms and $^3$He nuclei were polarized via spin-exchange collisions.

A distributed Bragg Reflector laser probed the potassium atoms via Faraday rotation. The probe laser was linearly polarized along the $y$ axis. The probe laser power was 3 mW and the wavelength was 0.5~nm red detuned from the $D_1$ transition in potassium at 770 nm. An electro-optic modulator and linear polarizer were used to stabilize the power of the probe laser. After passing through the vapor cell, the polarization of the probe laser light was monitored using a Wollaston prism and a balanced photodiode. The Faraday rotation signal is sensitive to potassium spin polarization pointing along the $x$-axis (see Figure \ref{fig:BlockDiagram}).

The optical setup and magnetic shields were housed in a plastic box that was insulated with 4 cm thick foam board insulation to improve thermal stability. The temperature inside the box was monitored with a DS18B20 digital temperature sensor. The ambient temperature in the box drifted at a level of $\approx 0.05\: ^\circ$C/hr. The magnetic field outside of the magnetic shields was measured with a 3-axis giant magnetoresistance magnetic sensor and was stable at a level of $\approx 100$~nT over the course of the many hours.

The signal from the balanced photodiode was filtered using a two-pole high-pass active Butterworth filter with a corner at 3.0 Hz and a two-pole low-pass active Butterworth filter with a corner at 16.0 Hz. The filtered signal was passed through an op-amp based phase shifting circuit and a variable gain amplifier. The amplified signal was used as the control voltage for voltage-controlled current supply. 

Once the $^3$He spins were polarized along the leading magnetic field by the pump beam, the nuclear spin oscillator could be started by tipping the spins of the $^3$He with a magnetic field pulse along the $y$ direction. After the $^3$He spins were tipped, feedback was used to keep the nuclear spin oscillator oscillating.  The oscillations of the $^3$He were detected by the probe laser as its polarization was rotated by the component of the magnetic field in the $x$ direction. This feedback allowed the nuclear spin oscillator to self sustain its oscillations.

The filtered oscillating signal from the balanced photodiode were recorded by a data acquisition box at a rate of 512 data points per second.

For the NSO to operate, the magnetic field inside the shields must be carefully controlled.  A leading magnetic field is applied along $z$, and residual fields along $B_x$ and $B_y$ or magnetic gradients were zeroed by applying magnetic fields with the coils in the shields.  The precession frequency of the atoms was used to zero $B_x$ and $B_y$ since it directly depends on the magnetic field inside the shields.  When $B_x$ and $B_y$ are zero, the overall magnetic field in the shields will be lower and the atoms will precess more slowly than when they are nonzero.  To zero $B_x$ and $B_y$, measurements of the precession frequency were made for a variety of magnitudes of small correcting magnetic fields applied by the coils. We estimate the transverse fields were zeroed to less than $\approx 10$~pT, making them negligible compared to the leading field of $\approx 300$~nT.    

The gradients were zeroed using the free induction decay time, $T_2^*$. The polarized spins were tipped by applying a magnetic field pulse in the $y$ direction and the decay of the precession was measured. The gradients were adjusted to maximize $T_2^*$. Through this method, decay times up to 1500 seconds were achieved.

\section{Results}

\begin{figure}[H]
\includegraphics[width=14.0 cm]{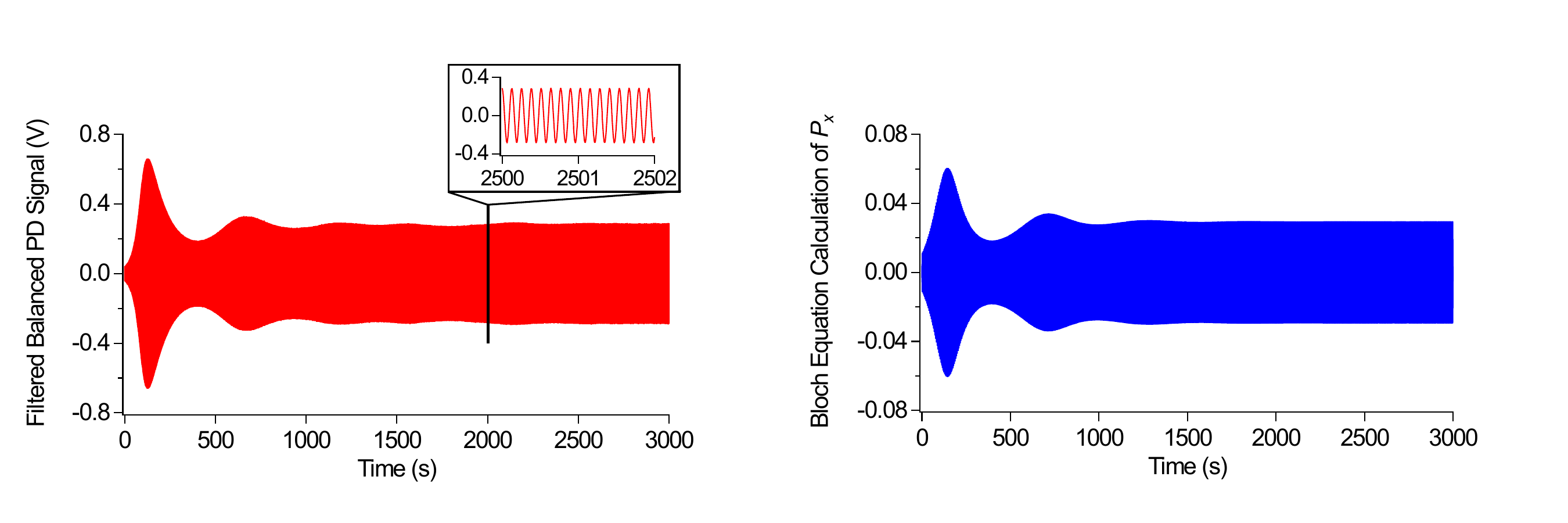}
\caption{Comparison of the experimentally observed start-up behavior of the nuclear spin oscillator (left) and the calculated behavior using the Bloch equations (right). The parameters used in the Bloch equations were adjusted to achieve qualitative agreement with the observed behavior.  \label{fig:NSOStartup}}
\end{figure}

The oscillator behavior during start-up can be compared with calculations from the model. While the nuclear spin oscillator can be started by initially tipping the spins, it can also spontaneously begin once the feedback is enabled due to slight misalignment of the pump laser beam with the leading magnetic field, stochastic fluctuations of the nuclear spins, or electronic or optical noise. Figure \ref{fig:NSOStartup} (a) shows this start up behavior and part (b) of the figure shows the solutions found by numerically solving the differential equations, Equations\ \eqref{eq:NSODiffEqA} - \eqref{eq:NSODiffEqC}. The parameters of the model, ($T_2$,  $\alpha$, $T_1^*$, and $P_0$ in the equations) were adjusted to qualitatively match the observed behavior.

\begin{figure}[H]
\includegraphics[width=14.0 cm]{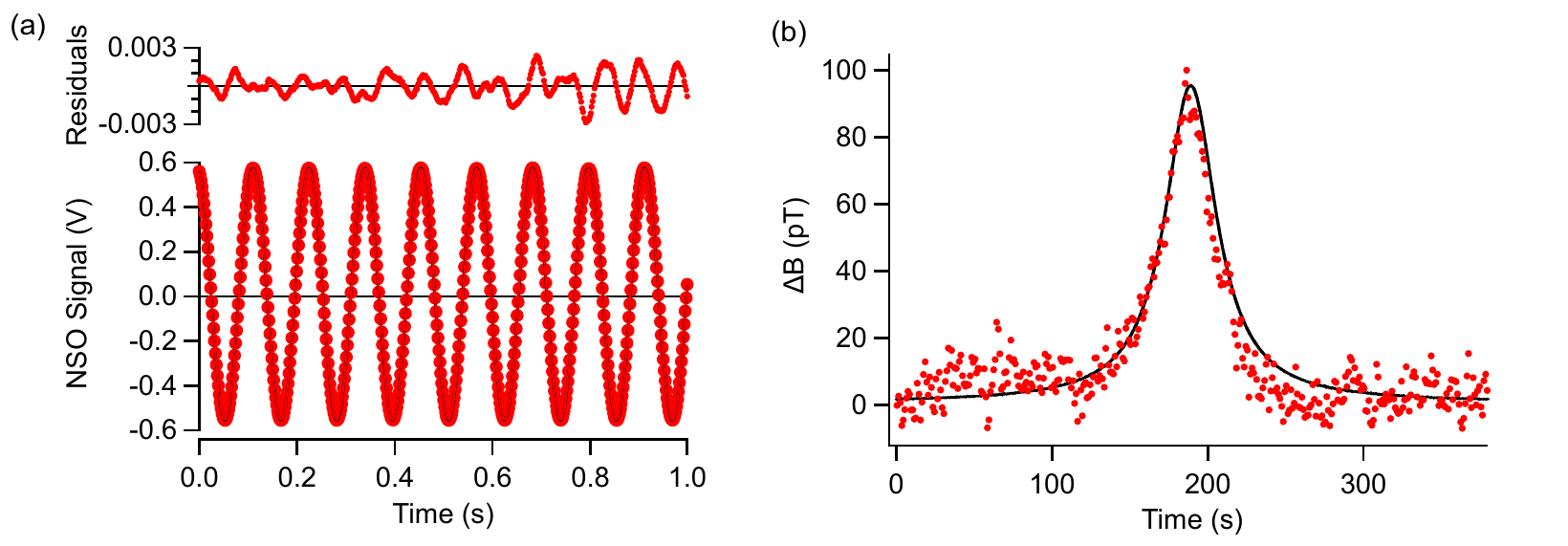}
\caption{Figure (a) shows the fit of the oscillator signal to a sine wave over a 1 second interval. The extracted frequency is proportional to the magnetic field in the $z$ direction. Figure (b) shows the response of the oscillator to an applied magnetic field in the $z$ direction. The frequency of the oscillator was determined at each second (red data points) and the frequency converted to a change in the magnetic field. The black curve shows the applied magnetic field. \label{fig:PulseFigure}}
\end{figure}

Following the start-up, the nuclear spin oscillator reaches steady-state oscillation with a constant oscillation amplitude. If the $z$ component of the magnetic field is changed, the frequency of the oscillation will change proportionally to the change in the magnetic field along the $z$ axis. This response was verified by applying a magnetic field pulse in the $z$ direction using one of the magnetic field coils within the magnetic shields. The magnetic field pulse was chosen to have a Lorentzian shape with a full-width-half-maximum of 40~s and a pulse amplitude of 100 pT. This is the type of signal that was searched for by the GNOME collaboration in Reference \cite{afach21}. The recorded oscillator data were binned in 1 second intervals and each interval was fit to a sinusoidal signal. An example of the fit is shown in Figure \ref{fig:PulseFigure} (a). The change in the magnetic field causes a change in the frequency of the oscillator. This frequency can be converted into a magnetic field using Equation \eqref{eq:FGammaB}. Figure \ref{fig:PulseFigure} (b) shows the change in the magnetic field measured by the nuclear spin oscillator as a function of time. The solid line in the figure shows the applied magnetic field. For these data the dc magnetic field was 273 nT, corresponding to an oscillation frequency of 8.87 Hz.

\begin{figure}[H]
\includegraphics[width=14.0 cm]{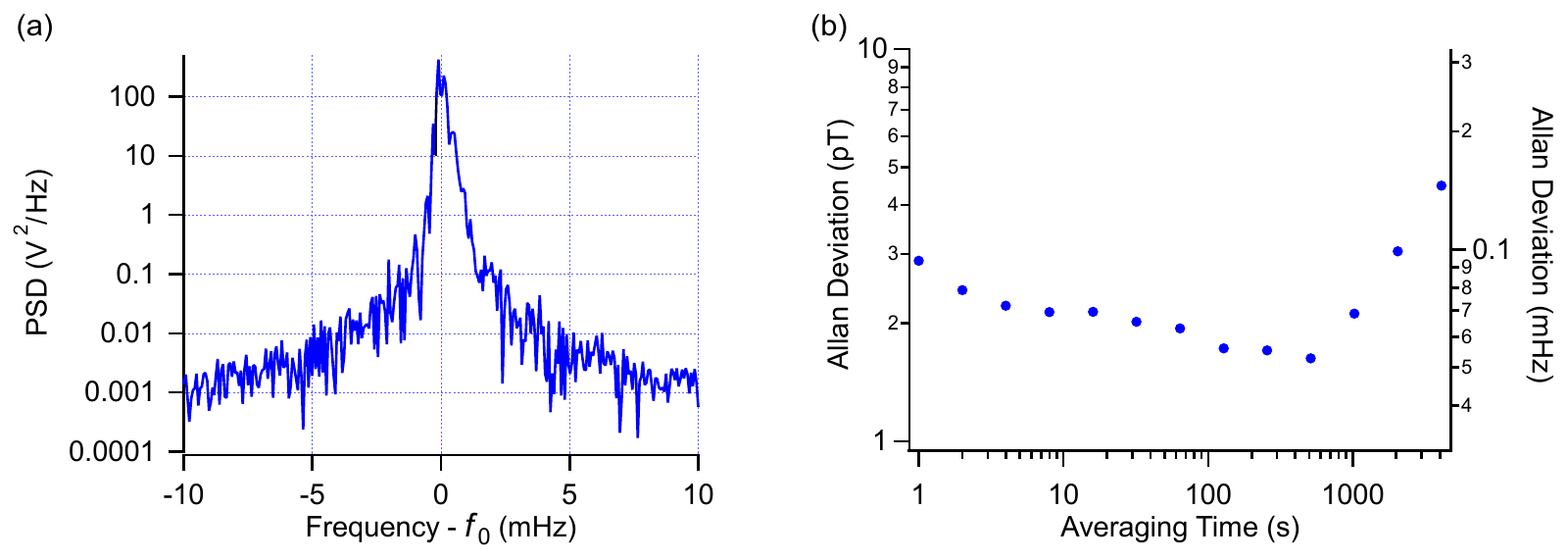}
\caption{Figure (a) shows the power spectral density (PSD) near the oscillator central frequency. The oscillator frequency of $f_0 = 8.8723$~Hz was subtracted from the frequency scale to more clearly show the line width of the oscillator. The signal is based on 4 hours of data. Figure (b) shows the Allan deviation of the oscillator in magnetic field (left axis) and frequency (right axis). The frequency at each second was determined by fitting one-second time intervals to a sinusoidal function. The oscillator improves with averaging time until $\approx 400$ s, at which point the frequency begins to drift. For these data the applied magnetic field was 273 nT, corresponding to a frequency of 8.87 Hz. \label{fig:PSDAllan}}
\end{figure}

To characterize the long-term behavior data were acquired for multiple hours. The power spectral density (PSD) of a four-hour section of data is shown in Figure \ref{fig:PSDAllan} (a). The central oscillation frequency of 8.8723 Hz was subtracted in the figure to more clearly show the width of the resonance, which we estimate to be $\approx 4$~mHz, FWHM. 

A four hour section of data were binned in one-second sections and fit to a sinusoidal signal, as was done with the pulse data described above. An Allan deviation calculated for the extracted frequencies is shown in Figure \ref{fig:PSDAllan}. The Allan deviation shows modest improvement with averaging time until times of $\approx 400$ s, at which point drift begins to be significant. The exact origin of the drift is unknown. However, the effect of ambient temperature on the magnetic field measurement was performed using a single-beam spin-exchange-relaxation-free magnetometer (SERF) that utilized the same magnetic shields \cite{park19}. In that work, the magnetic field measurement was observed to vary by $\approx 300$ pT/$^\circ$C. This drift was attributed to temperature variation in the shielding factor of the magnetic shields. Using this result, the measured temperature drift of 0.05 $^\circ$C/hr would lead to a drift in the frequency of the oscillator of $\approx 0.5$~mHz/hr. Other possible origins for the drift include variations in the gain or phase of the feedback electronics. 

An important aspect of magnetometers for a dark matter search is the bandwidth of the detector. The procedure of fitting the data in 1 second intervals limits the response to less than 1 Hz. This time interval was chosen to balance the bandwidth with signal with the robustness of the fit. Within this 1 second limitation, the response of the nuclear spin oscillator was evaluated by applying a square pulse magnetic field signal with amplitude of 100~pT and duration of 10 seconds. The nuclear spin oscillator response to the square pulse exhibited no delay or overshoot, indicating the bandwidth is limited by the fitting interval. More fundamentally, the frequency of the nuclear spin oscillator limits the bandwidth as a minimum of a full period is needed to obtain a reliable determination of the frequency.

\unskip

\section{Discussion}

The concept of using a magnetometer to search for exotic interactions relies on a spin-dependent coupling with a pseudo-magnetic field,  $\mathbf{\Upsilon}$. This pseudo-magnetic field may derive from a pseudoscalar dark matter field, $\phi$, via a gradient term $\mathbf{\Upsilon} = \boldsymbol{\nabla}{\phi}$. The interaction of this pseudo-magnetic field can then be characterized by an exotic dipole moment 
\begin{align}
  \hat{\boldsymbol{\chi}} = \chi \, \frac{\hat{\mathbf{F}}}{\abs{F}}, 
\end{align} 
where $\hat{\mathbf{F}}$ is the total angular momentum operator of the atom. The interaction is described by a Hamiltonian of the form \cite{afach23} 
\begin{align}
  \hat{\mathcal{H}} = & - \chi\, \frac{\hat{\mathbf{F}}}{\abs{F}} \cdot \mathbf{\Upsilon}  \\
   = & -\sum_{i=e,p,n} g_{\Upsilon\, i} \, \sigma_i\, \frac{\hat{\mathbf{F}}}{\abs{F}} \cdot \mathbf{\Upsilon} . 
\end{align}
Here, the sum is over the electrons, protons and neutrons in the atom, $\sigma_i$ is the spin content of the electrons, protons, and neutrons of the atom, and the $g_{i\, \Upsilon}$ are the couplings between the electron, proton, and neutron with the pseudoscalar field. For magnetically shielded magnetometers, the shielding material leads to an approximate cancellation of the interaction of the atomic spins with the pseudoscalar field \cite{jacksonkimball16}. As a result, the GNOME network is primarily sensitive to couplings to the nuclear spins. The alkali atom magnetometers used in the GNOME collaboration are most sensitive to the proton coupling, due to their nuclei having valance protons. For example, based on the Schmidt model of the nucleus, the exotic nuclear dipole moment of potassium is given by \cite{jacksonkimball15}
\begin{align}
  \chi\prn{^{39}\textrm{K}} \approx - 0.2 \, \chi_p. 
\end{align} 
In contrast, the exotic dipole moment of the $^3$He nucleus is largely dependent on a coupling to the neutron \cite{jacksonkimball15}
\begin{align}
  \chi\prn{^3\textrm{He}} \approx 0.87 \chi_n - 0.03\chi_p .
\end{align} 

While the couplings to different nucleons make a direct comparison of the two systems impossible, we can compare the detectors based on their energy sensitivity.  The nuclear spin oscillator described here has a sensitivity of $\delta B \approx 3$~pT in one-second of averaging. This corresponds to an energy sensitivity of 
\begin{align}
\delta E_\textrm{NSO}\prn{1 \: s} = \gamma \, \delta B \approx 4 \times 10^{-19} \: \textrm{eV}  
\end{align}
in 1 second of averaging. By comparison, an alkali magnetometer based on potassium that was used in the GNOME network, see e.g. Reference \cite{park19}, demonstrated a sensitivity of $\approx 0.1$~pT in 1 second of averaging. This corresponds to an energy sensitivity of 
\begin{align}
  \delta E_\textrm{K}\prn{1 \: s} = \frac{g_S\, \mu_B}{q} \, \delta B \approx 2 \times 10^{-18}\: \textrm{eV} ,
\end{align}
where  $\mu_B$ is the Bohr magneton, and $g_S \approx 2$ is the Land\'e factor for the electron, and $q = \frac{S\prn{S+1} +I\prn{I+1}}{S\prn{S+1}} = 6 $ is the nuclear slowing down factor in the limit of low alkali atom polarization \cite{allred02}, the regime in which this magnetometer operated.

We have evaluated a nuclear spin oscillator based on $^3$He nuclei. The system displays robust operation and has potential applications to dark matter searches. It offers modest improvement over the alkali atom magnetometers for the GNOME network in terms of both sensitivity and drift, albeit with a lower bandwidth. This may make it suitable for searches for some dark matter candidates, such as domain walls formed from axion-like particles that were searched for in Reference\ \cite{afach21}. We note that while the nuclear spin oscillator offers advantages over alkali atom magnetometers, the co-magnetometer technique developed in References \cite{kornack05,padniuk22,klinger23} have achieved energy sensitivities of $\approx 10^{-21}\: \textrm{eV}/\sqrt{\textrm{Hz}}$ for the neutron coupling and  $\approx 10^{-19}\: \textrm{eV}/\sqrt{\textrm{Hz}}$ for the proton coupling making them a much better system for developing advanced GNOME detectors. 




\vspace{6pt} 


\authorcontributions{Conceptualization, J.E.S.; methodology, H.R.P., A.M., S.C.W., S.T.Z. and J.E.S; formal analysis, H.R.P., A.M., S.C.W., S.T.Z. and J.E.S; investigation, H.R.P., A.M., S.C.W., S.T.Z. and J.E.S; writing---original draft preparation, J.E.S.; writing---review and editing, H.R.P., S.C.W, J.E.S.; supervision, J.E.S.; project administration, J.E.S.; funding acquisition, J.E.S. All authors have read and agreed to the published version of the manuscript.}

\funding{This research was funded by the U.S.\ National Science Foundation under grants PHY-1707803, PHY-2110370 and PHY-2510627.}

\acknowledgments{The authors acknowledge useful discussions with members of the GNOME collaboration, particular, D.F. Jackson Kimball and I.A. Sulai. The authors have reviewed and edited the output and take full responsibility for the content of this publication.}

\conflictsofinterest{The authors declare no conflicts of interest.} 



\abbreviations{Abbreviations}{
The following abbreviations are used in this manuscript:
\\

\noindent 
\begin{tabular}{@{}ll}
FWHM & Full Width at Half Maximum\\
GNOME & Global Network of Optical Magnetometers to Search for Exotic Physics \\
NSO & Nuclear Spin Oscillator \\
PSD & Power Spectral Density
\end{tabular}
}




\isPreprints{}{
\begin{adjustwidth}{-\extralength}{0cm}
} 

\reftitle{References}

\isPreprints{}{
\end{adjustwidth}
} 
\end{document}